# A room-temperature cavity-magnonic source of correlated microwave pairs

Qiuyuan Wang[1], Aravind Karthigeyan[1,2], Chung-Tao Chou[1,3], Luqiao Liu[1,*]

[1]Department of Electrical Engineering and Computer Science, Massachusetts Institute of Technology, Cambridge, MA, USA

[2]Department of Physics, University of Texas at Austin, Austin, Texas, USA

[3]Department of Physics, Massachusetts Institute of Technology, Cambridge, MA, USA

[*]Email: luqiao@mit.edu

**Abstract**

Correlated microwave photon sources are key enablers for technologies in quantum-limited sensing, signal amplification and communication, but the reliance on millikelvin operating temperature limits their scalability for broader applications. Here, at room temperature, we demonstrate strong correlated microwave signals emitted from a hybrid magnon-photon platform. Different from traditional parametrically induced magnons with degenerate frequencies, we achieve non-degenerate excitations by coupling magnon modes simultaneously with two cavity photon modes. Through the magnon-photon interactions in the corresponding linear and nonlinear regimes, one input microwave photon splits into a pair of magnon polaritons that possess distinct frequencies but maintain strong inter-mode correlations. The nonlinear magnon polariton dynamics empowered by this new parametric platform brings both verified true randomness and robust multi-channel correlations, from which we construct a microwave communication experiment for noise resilient signal transmission with added security. This work establishes cavity magnonics as a versatile and compact platform for generating correlated multi-mode microwave signals, opening new avenues for applications in classical and quantum domains.

In optics, when an incoming pumping photon is incident on a medium with strong second-order nonlinearity, spontaneous parametric down conversion (SPDC) or parametric oscillation can occur[1,2], splitting one pumping photon into a pair of new photons (Fig. 1c). The resultant interparticle correlation and entanglement have led to useful applications such as signal amplification, quantum key distribution, and intelligent imaging[3–10]. Today, parametrically generated photon pairs are typically realized using nonlinear crystals in the optical domain or Josephson junctions in the microwave domain. While these platforms have driven great technology breakthroughs[11–13], both face limits: nonlinear crystals allow room temperature operation, but it is difficult to realize two-qubit gates and bulky setups are required[14,15]; Josephson-junction-based quantum circuits enable tunable interactions but only works under cryogenic temperatures[8,16,17].

Here, we aim to develop a room temperature mechanism for generating correlated microwave photon pairs, where a hybrid magnon-microwave cavity system acts as a strong and tunable nonlinear medium. The magnonic system is known to exhibit rich nonlinear dynamics due to strong dipolar and exchange interactions among the quasiparticles. However, one critical restriction is that the parametrically generated magnons are in general degenerate[18,19] – the pair of magnons excited by an input microwave photon possess the same frequency and opposite wave vectors (Fig. 1h). The spectral indistinguishability and geometrical overlap of the two output modes make it challenging to study and utilize the correlation properties of such magnons for practical applications. To overcome this difficulty, we realize non-degenerate parametric down-conversion by coupling magnons with cavity-confined microwave photons (Fig. 1i). Benefiting from the gap opening from magnon-photon coupling and the nonlinear magnonic interactions, we excite magnon polariton pairs with split frequencies under a single microwave input. With the capability to discern the output signals as the *signal* and *idler* modes, we further reveal that strong correlations exist in the generated polariton pairs, featuring a room temperature correlated microwave photon source harnessing the quantum physics of the underlying cavity-magnonic system. Utilizing the uncertainty of individual polariton modes and the correlation between the two, we design and demonstrate a robust microwave communication experiment, where data bits can be effectively transmitted through noisy channels while keeping the information ciphered.

Our experiment and theoretical modelling address the interplay among magnons and multi-mode cavity photons in the existence of thermal baths, highlighting a new opportunity for studying nonlinear light-magnetic matter interactions, in complement to existing circuit or cavity quantum electrodynamics (cQED). The compact size, ambient working condition and high brightness also establish our device as a promising candidate for applications where correlated microwave photon pairs play a central role in enhancing sensitivity, rejecting noise and enabling quantum-limited amplification.

**Non-degenerate parametric pumping of magnon polaritons**

As a versatile quantum platform, cavity magnonics employs the hybridized dynamics between magnons and microwave photons to exchange energy and information coherently among different quantum sub-systems, for achieving useful functions including transduction, memory, sensing, and so forth[20–27]. To date, the majority of studies on cavity magnonics focus on the linear region where magnons interact with a single cavity-photon mode[28-30]. Here, to investigate spontaneous generation of quasiparticles through nonlinear dynamics, we design and implement a hybrid system where magnons interact simultaneously with two cavity modes which collectively account for the formation of parametric magnon polaritons.

Fig. 1a shows the device structure, where an Yttrium iron garnet (YIG) film with an area of $2.0 \times 1.1$ mm$^2$ and thickness of 3 μm is mounted onto a two-port microstrip resonator with the magnetic film facing down. In order to maximize the coupling strength between the resonator modes and magnons, and at the same time maintain a reasonable quality factor (Q), the microstrip is designed to maintain a large channel width (0.89 mm) over the majority of its length and narrow down to 40 μm for a short segment where the magnetic sample is located (see Fig. 1a). We fine tune the size and shape of this capacitively-coupled microstrip resonator, such that the two resonance modes with the lowest frequencies – half wavelength ($\frac{1}{2}\lambda$) mode at $\omega_{r1}$ and full wavelength ($\lambda$) mode at $\omega_{r2}$, satisfy the exact frequency doubling relationship $\omega_{r2} = 2\omega_{r1}$ (Supplementary Note 3), which in general is not guaranteed in these loaded resonators[31]. The transmission spectrum S$_{21}$ of the device is shown in Fig. 1e, where $\frac{\omega_{r1}}{2\pi}$ and $\frac{\omega_{r2}}{2\pi}$ are precisely aligned at 3.354 GHz and 6.709 GHz, respectively.

For parallel pumping of magnons, a radio frequency (rf) magnetic field at twice the magnon frequency (i.e., $\omega_{r2}$) needs to be applied parallel to the equilibrium direction of the magnetic moment $\boldsymbol{m}$. Meanwhile, a rf field at $\omega_{r1}$ that is perpendicular to $\boldsymbol{m}$ is optimal for maximizing the magnon-resonator coupling strength in the linear regime. In order to satisfy both requirements simultaneously, we apply the static magnetic field $H_0$ in the film plane with a ~ 38° angle to the in-plane rf field direction generated from the microstrip signal line (Fig. 1a) and position the YIG sample at ~$\frac{1}{4}\lambda$ location (Fig. 1d). Figure 1f shows the S$_{21}$ spectrum measured in the linear domain as a function of $H_0$. We see anti-crossings form at $\omega_{r1}$ and $\omega_{r2}$ (Fig. 1g), with corresponding magnon-photon coupling strengths[20] (for Kittel mode, wave vector $\boldsymbol{k} = 0$) fitted to be $\frac{g_1(\boldsymbol{k}=0)}{2\pi} \approx$ 70.3 MHz and $\frac{g_2(\boldsymbol{k}=0)}{2\pi} \approx$ 61.9 MHz. The dissipation rates of the two resonator modes and the ferromagnetic resonance (FMR) magnon around $\omega_{r1}$ are $\frac{\kappa_{r1}}{2\pi} \approx$ 82.4 MHz, $\frac{\kappa_{r2}}{2\pi} \approx$ 144.4 MHz, and $\frac{\kappa_{m1}}{2\pi} \approx$ 68.8 MHz (fitting details in Methods).

We study the parametric magnon polaritons by applying a continuous pumping microwave signal at $\frac{\omega_p}{2\pi} = 6.708$ GHz with a 12 dBm power (at the output of signal generator, corresponding to a peak field strength estimated to be $h_{\rm rf} \sim 38$ Oe at the sample location, see Methods) from the left port of the resonator and measuring the power spectrum of the output signals from the right port (Fig. 1b). The dc magnetic field $H_0$ is swept around the corresponding FMR condition for $\frac{\omega_p}{2}$. Figure 2a and 2b show the measured power spectra as a function of applied $H_0$ and the frequency readout from a spectrum analyzer in the vicinity of $\frac{\omega_p}{2}$. We found that as $H_0$ increases, a single sharp peak located at $\frac{1}{2}\omega_p$ first emerges, corresponding to standard degenerate parametric pumping. With the increase of $H_0$, a pair of non-degenerate output modes emerge from the background, where the two peaks appear symmetrically around $\frac{\omega_p}{2}$, and the detuning from the central frequency $\Delta = \omega - \frac{\omega_p}{2}$ increases with $H_0$.

To verify the nonlinear nature of the observed signals, we measured the power dependency of the output spectra (Fig. 2c). For both degenerate and non-degenerate modes, the existence of threshold power and abrupt increases of output signal above the threshold are consistent with the picture of a nonlinear, parametric process. It is worth noting that under the same magnetic field, tuning the power of the driving signal does not toggle the system between degenerate and non-degenerate modes, or change the frequency splitting and linewidths in the non-degenerate case. The relatively robust frequency splitting suggests that the output microwave photons are from intrinsic modes of the cavity-magnonic system, rather than other potential mechanisms like microwave dressed states[32], where the frequency difference is expected to increase linearly with the driving rf field.

Next, we provide a unified model covering the parametric excitation of both degenerate and non-degenerate modes by considering a minimum Hamiltonian including a single pair of magnon modes $\hat{c}_{\pm k}$ with opposite wave vector $\pm k$, the resonator mode $\hat{a}_{r1}$ at $\omega_{r1}$, and the pumping microwave field $\hat{a}_{r2}$ at $\omega_{r2}$, where $\hat{c}_{\pm k}$, $\hat{a}_{r1}$ and $\hat{a}_{r2}$ are annihilation operators:

$$\frac{\hat{H}}{\hbar} = \omega_{\pm k}\hat{c}^\dagger_{\pm k}\hat{c}_{\pm k} + \omega_{r1}\hat{a}^\dagger_{r1}\hat{a}_{r1} + \omega_{r2}\hat{a}^\dagger_{r2}\hat{a}_{r2} + g_{\pm k}(\hat{a}^\dagger_{r1}\hat{c}_{\pm k} + \hat{c}^\dagger_{\pm k}\hat{a}_{r1})$$
$$+ P_k(\hat{a}^\dagger_{r2}\hat{c}_k\hat{c}_{-k} + \hat{c}^\dagger_k\hat{c}^\dagger_{-k}\hat{a}_{r2}) + \frac{\hat{H}_{4m}}{\hbar} \qquad (1)$$

Here, $g_{\pm k}$ and $P_k$ reflect the coupling strengths of corresponding resonator modes. $\hat{H}_{4m}$ represents the four-magnon scattering that accounts for the Kerr effect and nonlinear damping. Under zero detuning between magnon and $\omega_{r1}$ mode ($\omega_{\pm k} = \omega_{r1} = \frac{\omega_p}{2}$) and taking into account magnon and resonator dissipation, the excited parametric magnon polaritons have eigenfrequencies of $\omega_\pm \approx \frac{\omega_p}{2} \pm \sqrt{2g_k^2 - \kappa^2}$ ($\kappa$ is the weighted average dissipation rate for the photon and magnon modes at $\omega_{r1}$; Supplementary Note 1). This result explains the observed evolution from degenerate to non-degenerate excitations with the increase of $H_0$, as well as the small frequency splitting in the non-degenerate case. When $H_0$ is low, the resonator mode crosses the magnon band at large wave vectors $k$ (Fig. 2c, middle inset). Meanwhile the coupling strength $g_k$ has a dependence on $k$ such that a higher $k$ yields a lower $g_k$ due to the internal cancellation within a magnon wavelength (Supplementary Note 2). For $2g_k^2 < \kappa^2$, $\omega_\pm$ has no splitting in its real part, corresponding to the degenerate excitation. With the increase of external field $H_0$, the bottom of magnon band shifts upward and the (anti-)crossing points move towards $k = 0$ and $g_k$ increases, leading to finite frequency splitting, in agreement with the trend in Fig. 2b. The expression of $\omega_\pm$ is also consistent with the negligible power dependence in Fig. 2c as the formula does not explicitly contain power, a result of the self-limiting effect on the oscillation amplitude through $\hat{H}_{4m}$ (details in Supplementary Note 1).

**Correlation of parametric magnon polaritons**

When a single input photon excites a pair of magnons or magnon polaritons through the parametric process, the two generated quasiparticles are correlated, corresponding to continuous variable entanglement in the single or few quanta limit[33]. Notably, the parametric excitation term in Eq. (1), $P_k(\hat{a}_{r2}^\dagger \hat{c}_k \hat{c}_{-k} + \hat{c}_k^\dagger \hat{c}_{-k}^\dagger \hat{a}_{r2})$ corresponds to a two-mode squeezing operator when treating the $\omega_{r2}$ driving mode classically, and the Hamiltonian for parametric magnon polariton has the same form as in optical SPDC or a non-degenerate Josephson parametric amplifier[34,35]. In our room temperature experiment, although the fluctuations in the two modes are no longer quantum-limited, a strong correlation between the output signals is still maintained. Theoretically we show that in the steady state the correlation function $\langle \hat{p}_+(t)\hat{p}_-(t)\rangle$ has a constant phase even when considering thermal noise, ensuring a phase anti-correlation between the two modes. Here $\hat{p}_+$ and $\hat{p}_-$ are the annihilation operators of the dominant magnon polaritons with non-degenerate frequencies (Supplementary Note 1).

We investigate the output microwave signals in the time domain to reveal the connection between the two non-degenerate modes. Figure 3a illustrates a typical output waveform when inspected under an oscilloscope after down-converting the central frequency from $\frac{1}{2}\frac{\omega_p}{2\pi} = 3.354$ GHz to 66.6 MHz. The sinusoidal waveform confirms the coherence of output signals on a short time scale, with the beating feature in the amplitude implying the existence of more than a single frequency but with fixed phase relationship. To characterize the output signals over a longer period of time, we measure the quadrature components[36] $(X_1, P_1)$ and $(X_2, P_2)$ of the two modes simultaneously using homodyne detection methods by splitting and down-converting the two signals with respect to their corresponding central frequencies (Extended Fig. 1). Figure 3b illustrates both the short-term and long-term evolutions of the measured quadrature signals. The quadrature components of each individual mode which remain unchanged within a few hundred nanoseconds as suggested in Fig. 3a, however, show large fluctuations on a longer time scale. The long-term randomness is also reflected in the two-dimensional distribution plots of quadrature components (Fig. 3c), where points corresponding to $(X_{1(2)}, P_{1(2)})$ stay on a circle with the phase evenly distributed from 0 to $2\pi$. The randomly distributed phase through non-degenerate pumping is drastically different from the case in degenerate pumping, where the phase of the output signal is fixed at 0 or $\pi$ (see Fig. 3d, measured from the same device). This fixed phase in the degenerate parametric pumping is known before, as a consequence of locking to the pumping input signal[37,38].

While individual outputs behave stochastically, when combined together the two signals exhibit deterministic nature in their superposition. The bottom panels of Fig. 3c demonstrate that the in-phase components ($X_1$ and $X_2$) are correlated and the out-of-phase components ($P_1$ and $P_2$) are anti-correlated, a signature of opposite phases between the two signals. This anti-correlated phase relationship is also summarized in the plots of $\varphi_1 \pm \varphi_2$, in the bottom panels of Fig. 3b and the histograms of Fig. 3f, where $\varphi_1 + \varphi_2$ is narrowly centered around zero and $\varphi_1 - \varphi_2$ evenly distributes over all possible values. We quantitatively characterize the statistical properties of the signals by calculating the correlation coefficients in Fig. 3g, $\rho_{A_i B_j}(\tau) = \frac{\langle A_i(t) B_j(t+\tau) \rangle}{\langle A_i \rangle \langle B_j \rangle}$ where $A(B) = X, P$ and $i(j) = 1,2$. Specifically, the correlation coefficient matrix for zero time delay ($\tau = 0$) is shown in Fig. 3e, where a large cross-correlation coefficient of $\pm 0.98$ is reached, confirming the strong correlation between the signals. Fitting Fig. 3g with $\rho_{X_i Y_j}(t) \propto e^{-t/\tau_c}$, we determine a coherence time $\tau_c = 0.18$ ms for both auto- and cross-correlations, in reasonable agreement with the oscillation linewidth in the spectrum measurement of Fig. 2, $\frac{\Delta\omega_{osc}}{2\pi} \approx$ 6.7 kHz, corresponding to $\tau_c = 0.15$ ms. We note that this coherence time is also consistent with the theoretically calculated value $\tau_c = 4\bar{n}/(\bar{n}_{th}\kappa) \approx 0.12$ ms when considering a thermally driven diffusive oscillator model evaluated using the thermal magnon polariton number $\bar{n}_{th}$, dissipation rate $\kappa$ and excited magnon polariton population $\bar{n}$ determined from the radial distribution of Fig. 3c (Supplementary Note 4).

The consistency between the measured and theoretical coherence time confirms that the fluctuations of individual oscillations' phase originate from thermal noise, as illustrated in the quantum Langevin equation of the system (Supplementary Note 1). Similar to the case where uncertainty is seeded by the quantum vacuum noise[39,40], the signals inheriting the thermal noise source have a true randomness nature as well. In Fig. 3h, we sample the phase from one of the output signals with different intervals, followed by standard digitizing processes and examinations against the NIST Statistical Test Suite[41]. We see that the generated signal passes all available randomness test criteria when the sampling interval is much greater than $\tau_c$.

**Microwave communication using correlated polariton pairs**

Correlated photon pairs from parametric down-conversion processes have become the key enabler of a number of emerging technologies, such as quantum communication[5], quantum radar[8] and adaptive imaging[10], where the cross-correlation between the signal and idler photons is harnessed for enhancing security, rejecting environmental noise and improving resolution. The combination of true randomness and phase correlation in the non-degenerate magnon polariton discovered in this work can be used for similar applications, with the added advantage of compact size and room temperature performance. As an initial demonstration, we present a microwave communication experiment utilizing the strongly correlated *signal* and *idler* pair for robust data transmission. The *signal* frequency is broadcast in the public domain as information carrier, while the *idler* is stored locally or transmitted privately for correlation operation at the receiver end (Fig. 4a). Employing the experimental setup of Fig. 4b, we first generate the correlated microwave pair with $\frac{\omega_1}{2\pi} = 3.3532$ GHz (*signal*) and $\frac{\omega_2}{2\pi} = 3.3548$ GHz (*idler*) through non-degenerate parametric pumping. After separating the two frequencies, we encode the information by phase-modulating the *signal* channel through quadrature phase-shifted keying (QPSK) under a 500 Hz modulation frequency. The modulated *signal*, after being transmitted through a transmitter antenna and picked up by a receiver antenna, is then mixed with the *idler* for data decoding.

As shown in the top panels of Fig. 4c, the phases of individual channels remain random during this communication process. Meanwhile, after combining the two using the fact of $\varphi_1 + \varphi_2 = 0$, we decode the transmitted information (bottom panel of Fig. 4c and the histogram in Fig. 4d). Figure 4e illustrates an example of transmitting a binary picture with 475 pixels, where the original image (left figure) is successfully reconstructed with no bit-error after the transmission (right figure) by utilizing the correlation between two noisy channels (two middle sub-figures).

**Discussion**

Correlated photon or polariton pairs are jamming-resilient because uncorrelated noise can be efficiently filtered out via correlation between the two channels at the receiver side, significantly improving the signal-to-noise ratio (SNR), which is particularly useful for communications in hostile environments (Fig. 4a; Supplementary Note 6). Similar benefit is also expected in other application areas such as radar where correlation improves the noise rejection capability[42]. With further suppression of thermal contributions by lowering the device working temperature, we expect to achieve unconditional security, where quantum entanglement between the two modes can be established and detected. There the expected two-mode squeezed state could also provide a platform for reaching quantum-limited magnetic field sensing.

To conclude, we present an engineered YIG/microstrip resonator device as a room-temperature source of correlated microwave signals. The coupling between magnons and resonator photons opens a frequency gap and enables non-degenerate parametric pumping of magnon-polariton pairs. Thanks to the nonlinear magnon dynamics and the synergetic interactions between magnons and photons, the generated two magnon polariton modes show strong phase anti-correlation with a moderate coherence time. In addition to revealing a new parametric magnon polariton physics, we prototype a microwave communication experiment that could be potentially used in hostile environments, combining the true randomness from individual signals with the cross-correlation between two channels. This compact, highly tunable, correlated microwave source opens doors for advanced applications in both the classical and quantum domains including hardware-based secure communication systems and quantum networks with outstanding versatility and scalability.

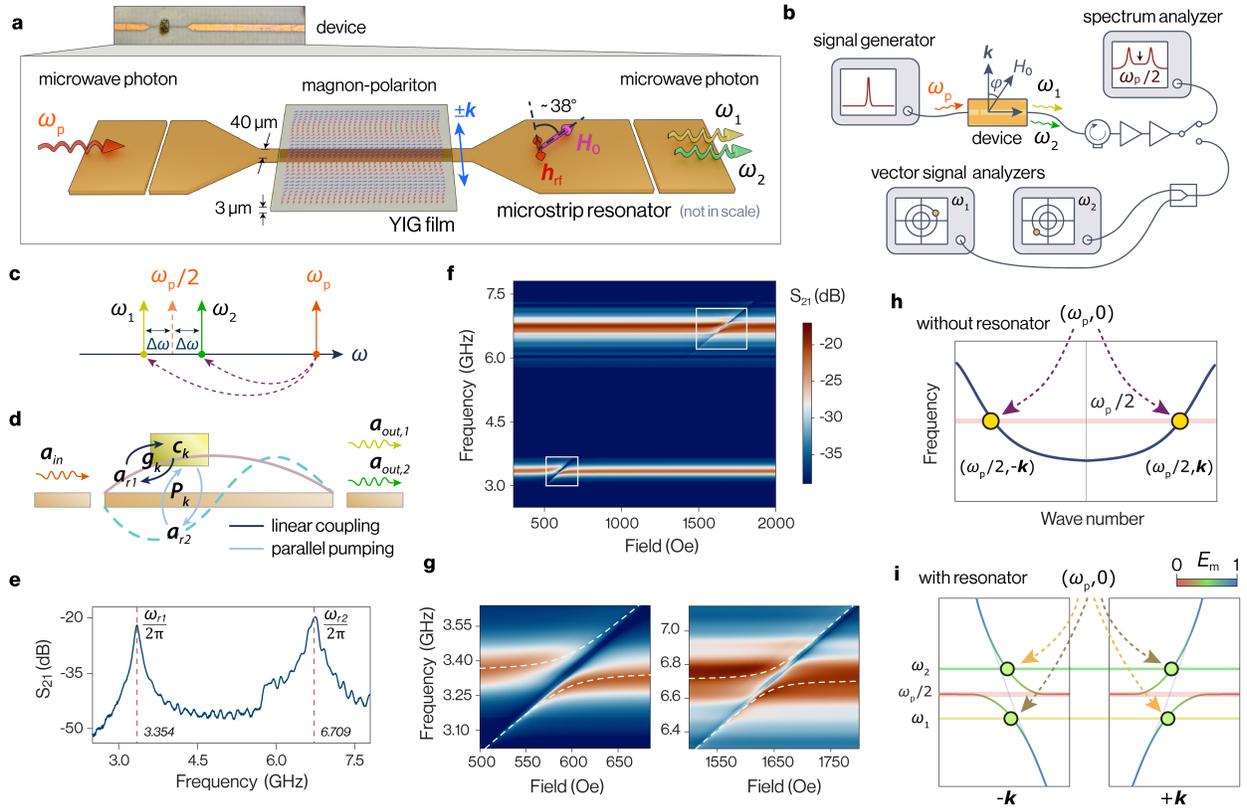

**Fig. 1 | Coupled YIG/resonator device for non-degenerate parametric pumping. a,** Schematic of the device. A piece of YIG film (3 μm thickness) is mounted onto a microstrip resonator. A continuous wave microwave signal with angular frequency $\omega_p$ is fed from the left port as pump. The output microwave fields ($\omega_1$ and $\omega_2$) are measured at the right port. $H_0$ is the static external magnetic field; $h_{rf}$ is the a.c. magnetic field from the pump. The excited magnon-photon polaritons consist of magnon modes propagating along the $\pm k$ direction. The schematic is not drawn in scale for showing the device details. **b,** Experimental setup for spectrum and quadrature measurements. $\varphi$, angle between magnon wave vector $k$ and external field $H_0$; $H_0$ is always applied in-plane. **c,** Energy diagram of the non-degenerate parametric pumping process. An input photon ($\omega_p$) is converted to a pair of magnon polaritons ($\omega_1$ and $\omega_2$). **d,** Illustration of a minimal model showing the interactions of magnons with two cavity photon modes. ($a_{in}$, input waveguide microwave field;

$a_{r1(2)}$, intracavity microwave field of the first and second resonant modes; $c_k$, magnon mode with wave vector $k$; $g_k$, coupling strength between $c_k$ and $a_{r1}$; $P_k$, coupling strength between $c_k$ and $a_{r2}$ under parametric process; $a_{\text{out},1(2)}$, output waveguide microwave field.) **e,** Transmission curve of the resonator. The sample is loaded with YIG film under -20 dBm input power and no external field. **f,** Measured transmission spectra $S_{21}$ as a function of field and frequency under -20 dBm input power. **g,** Zoomed-in plots around the anti-crossing regions in (**f**) for $\frac{\omega_{r,1}}{2\pi}$ and $\frac{\omega_{r,2}}{2\pi}$. The dashed lines represent fitting from input-output theory of a two-port resonator loaded with a magnet. **h,** Illustration of standard magnon parametric pumping process without coupling to resonator. The magnon dispersion curves are calculated for a 3 μm thickness YIG film under 600 Oe external magnetic field, and magnons propagating along $\varphi = 38°$ direction. **i,** Illustration of the non-degenerate parametric pumping process when the YIG film is coupled with a resonator that supports a $\omega_p/2$ resonant mode. The original $\omega_p/2$ magnons split into pairs of magnon-polariton modes. $E_m$: magnon proportion in a corresponding magnon-photon polariton.

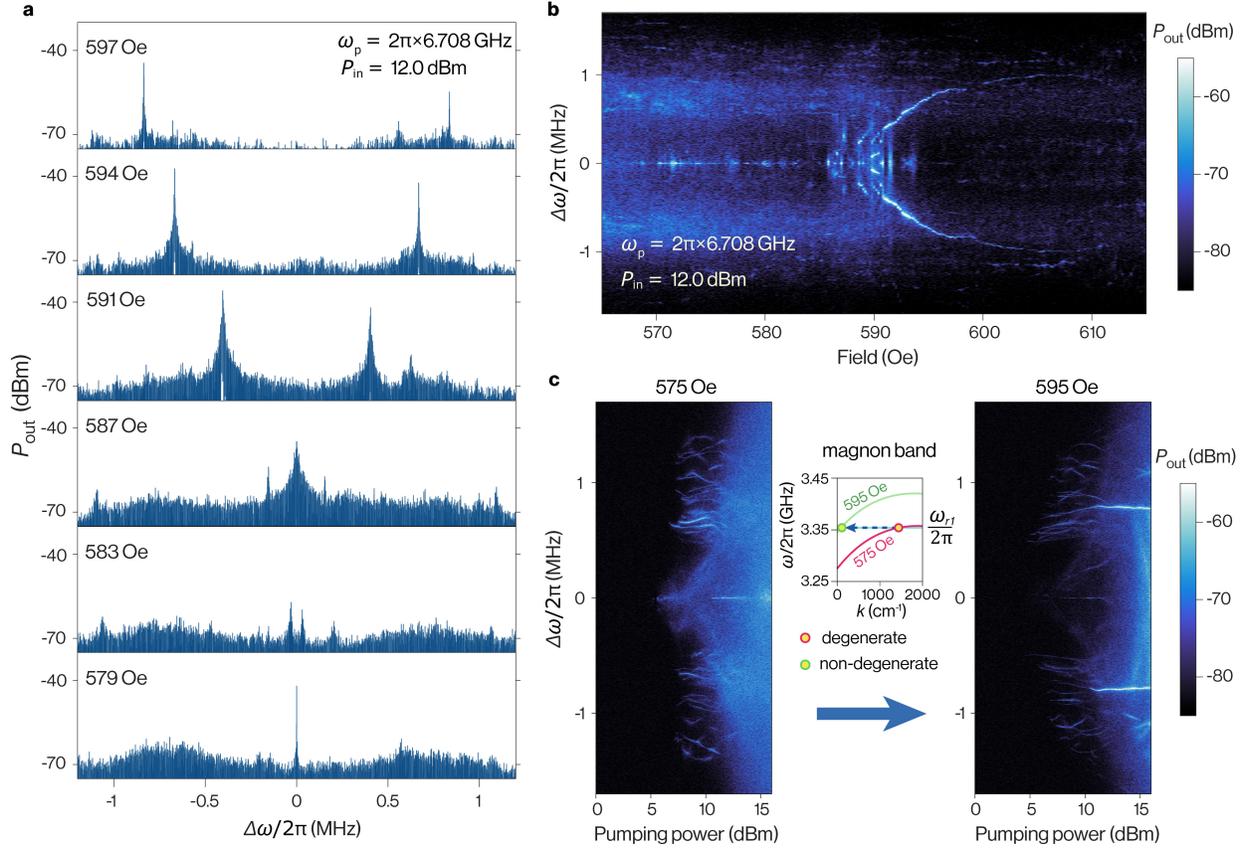

**Fig. 2 | Field and power dependency of degenerate and non-degenerate parametric pumping.**
**a,** Representative output power spectra around ω$_p$/2 under different magnetic fields and 12.0 dBm input power. $\Delta \omega = \omega - \frac{\omega_p}{2}$ is the detuning. **b,** Output power spectra as a function of magnetic field under 12.0 dBm input power. A pair of non-degenerate parametric pumping peaks is seen above ~590 Oe. **c,** Representative output power spectra as a function of input pumping power for degenerate parametric pumping when $H_0$ = 575 Oe (left) and non-degenerate pumping when $H_0$ = 595 Oe (right). A power threshold and saturation of signal peaks are observed for both degenerate and non-degenerate modes. The frequency splitting of the non-degenerate modes is robust against pumping power. The pumping signal frequency is fixed at 6.708 GHz. The middle inset illustrates the change of magnons' wave vectors ***k*** at the crossing point with the resonator $\omega_{r1}$ mode as $H_0$ increases. Lower ***k*** corresponds to higher $g_k$.

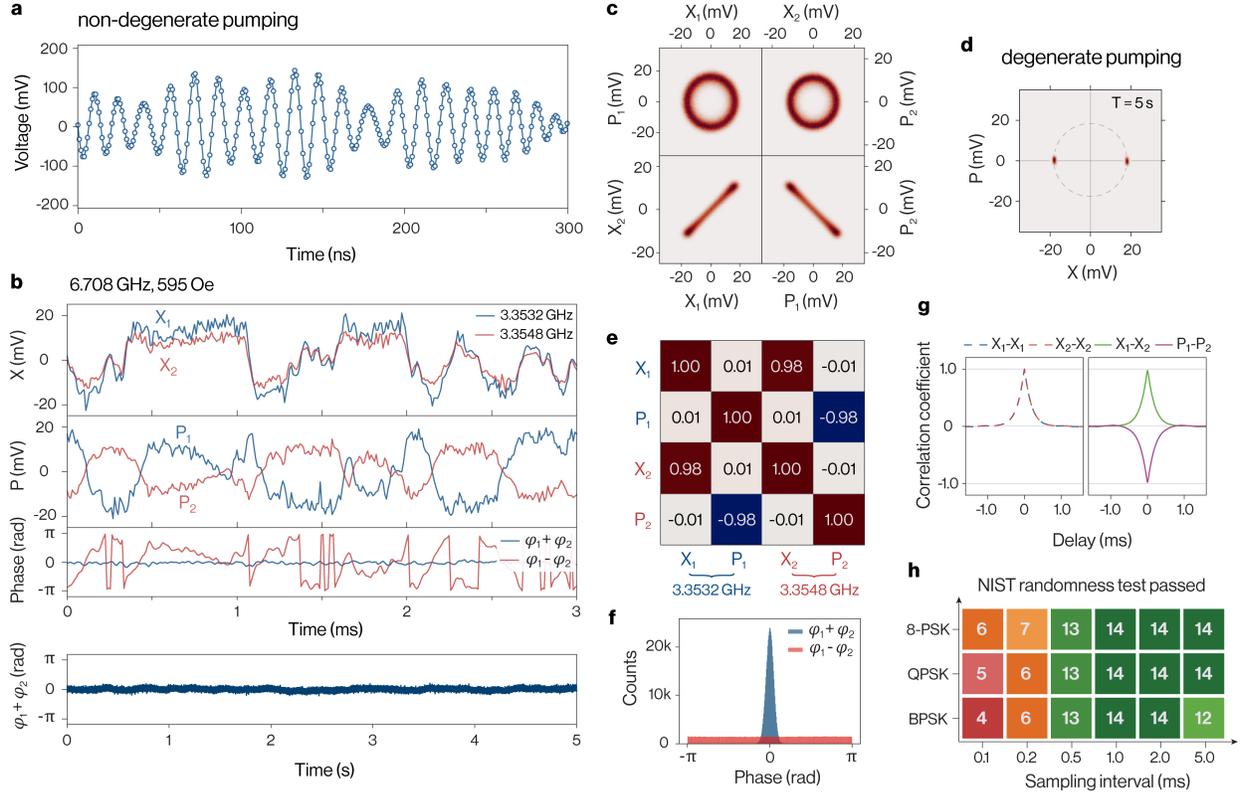

**Fig. 3 | Magnon polariton dynamics and correlation. a,** Example of output signal waveform (down-converted to 66.6 MHz) from non-degenerate parametric pumping. The voltage is measured after amplification. **b,** Measured time evolution of the in-phase (X), quadrature (P) components, phase sum ($\varphi_1 + \varphi_2$) and difference ($\varphi_1 - \varphi_2$) of a pair of non-degenerate outputs peaked at 3.3532 GHz and 3.3548 GHz. The lower panel shows evolution of phase sum $\varphi_1 + \varphi_2$ over a longer (5 second) period. **c,d,** Distribution of the output quadrature components for non-degenerate (**c**) and degenerate (**d**) parametric pumping. **e,** Cross-correlation coefficient matrix of the output quadrature components. Diagonal blocks represent autocorrelation. **f,** Distribution of phase sum and phase difference of the two output signals. **g,** Measured correlation coefficients as a function of delay time. **h,** Results from the NIST randomness test. Bitstreams are generated from one of the measured quadrature waveform (3.3532 GHz) using different digitizing coding schemes (BPSK: binary phase-shift keying; QPSK: quadrature phase shift keying; 8-PSK: 8- phase shift keying)

and sampling intervals from 0.1 ms to 5.0 ms. Numbers indicate the tests passed out of a total of 14. Maurer's Universal Statistical Test and Approximate Entropy Test are excluded due to insufficient bitstream length, a known limitation as both tests require data from extremely long acquisition time, challenging under current sampling rate. All measurements in this figure are performed under 6.708 GHz pumping signal with 15.0 dBm input power and a 595 Oe magnetic field.

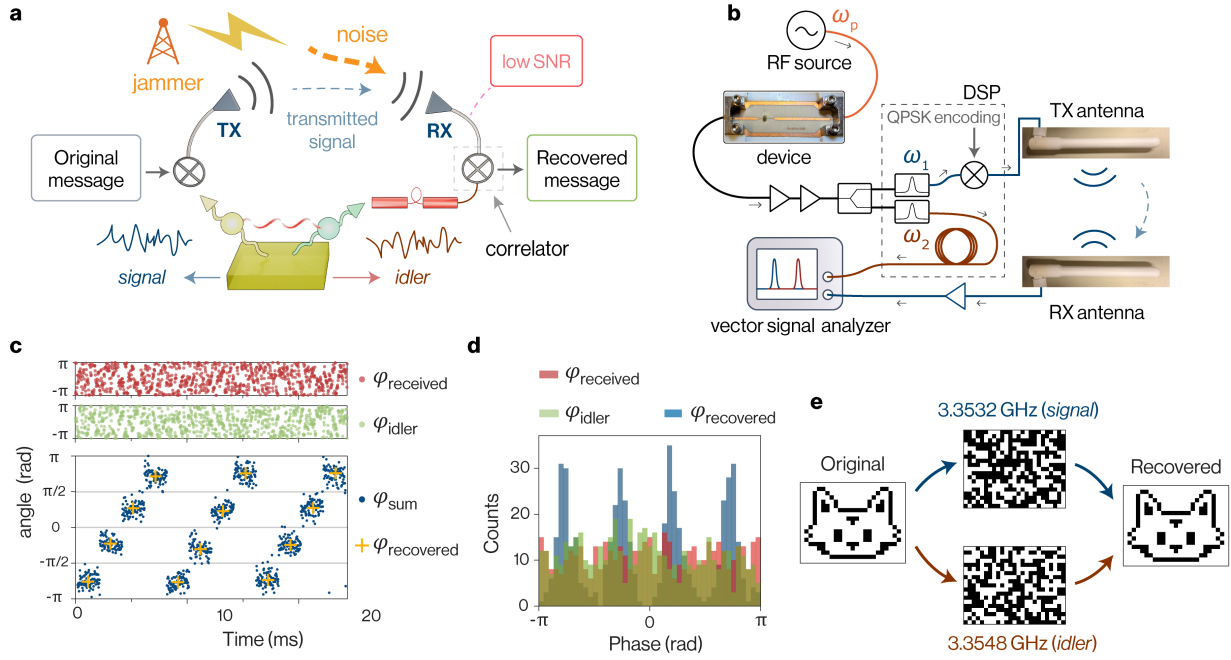

**Fig. 4 | Wireless communication experiment with correlated magnon polaritons. a,** Schematic of the communication experiment. The generated correlated microwave *signal* and *idler* pair is separated into two channels – one for public signal broadcasting, and the other is stored locally or transmitted privately for correlation operation at the receiver. During signal transmission, unwanted noises from the environment or a jammer may significantly reduce the SNR. The message can be effectively recovered by correlating the received noisy signal with idler. **b,** Experimental setup. The channel splitting and data encoding are performed by digital signal processing (DSP) using two vector signal transceivers (VST). The modulated signal is transmitted and collected through a pair of RF antennas. The idler and received signal are sent to the VST for correlation calculation. **c,** Example of the measured phase data. Upper panel: measured phase of received signal $\varphi_{\text{received}}$ and the idler $\varphi_{\text{idler}}$. Lower panel: phase sum of the two signals. The yellow cross symbol represents average within a 2-ms phase modulation window. The data is transmitted through QPSK keying, with quadrature modulation and sampling rate on the transmit

(TX) and receive (RX) side being 100 kHz and 800 kHz, respectively. **d,** Histogram of the signal and idler phases, as well as their sum after a uniform QPSK phase modulation shown in **c**. **e,** Demonstration of image transmission. The middle panels show transmitted data when monitored and decoded only using individual channels; the right panel shows the recovered image combining the signal and idler channels. All data in **c-e** are collected under 15.0 dBm continuous wave pumping signal at 6.708 GHz and a 595 Oe field.

## Methods

### Device fabrication

The microstrip resonator was made on double-sided Rogers RO4003C laminate with 0.41 mm dielectric thickness and 17.5 μm copper layer. The resonator pattern was defined using an LPKF ProtoLaser U4 laser cutter. The YIG film was grown on a piece of GGG substrate using liquid-phase epitaxy. It was then cut into pieces and mounted onto the resonator using GE varnish. The microstrip resonator was designed based on simulation results from Sonnet for impedance and frequency matching. The resonator parameters are fine-tuned based on experiment results. The peak value of RF magnetic field strength from the pumping microwave signal at YIG film surface is estimated using $h_{\rm rf,P} \approx \frac{1}{w}\sqrt{\frac{1}{T_1}\frac{\kappa_e}{\kappa_e+\kappa_c}\frac{P_{in}}{Z_0}} = \frac{1}{w}\sqrt{\frac{\omega_{r2}}{\kappa_e+\kappa_c}\frac{P_{in}}{Z_0}} = \frac{1}{w}\sqrt{\frac{QP_{in}}{Z_0}}$, where $T_1$ is the power transmission for one port at $\omega_{r,2}$; $\kappa_e$ is the total external dissipation rate; $\kappa_c$ is the intrinsic dissipation rate of the resonator; $Q \approx 46.5$ is the resonator quality factor for $\omega_{r,2}$ mode; $Z_0 = 50\,\Omega$ is the characteristic impedance of the microstrip transmission line; $w = 40$ μm is the microstrip width beneath YIG film; $P_{in}$ is the input pumping power. Transmission line attenuation is estimated to be ~0.3 dB and is therefore neglected.

### Measurement setup

The microwave circuits used for different types of measurement are shown in Extended Figure 1. The input pumping signal is generated using an Anritsu 68369A signal generator. The output signal from the device is sent through an isolator (Fairview SFC 2040A), a low-pass filter (Mini-Circuits VLFG-3500+), two low-noise amplifiers (Mini-Circuits ZX60-63GLN+, each with 24.5 dB gain), and measured using a spectrum analyzer (Anritsu MS8609A), oscilloscope (Siglent SDS 1104X-E) or two vector signal analyzers (National Instruments PXIe-5644R). For quadrature measurement, the frequencies of the non-degenerate modes are first extracted from output power spectrum (directly measured from the vector signal analyzers); then, the extracted frequencies are used for quadrature measurements. All microwave equipment is synchronized with a 10-MHz frequency standard.

**Magnon dispersion of YIG film**

The equations below[43] are used to calculate the dispersion of pure magnon frequencies in an extended YIG film with finite thickness $t$:

$$\omega_n^2 = (\omega_H + \omega_M \lambda^2 k_n^2)(\omega_H + \omega_M \lambda^2 k_n^2 + \omega_M F_{nn})$$

where $\omega_H = \gamma H$, $\omega_M = \gamma \cdot 4\pi M_S$, $\lambda = \sqrt{\frac{D}{4\pi M_S}}$, $D = \frac{2JSa^2}{\gamma \hbar}$, $k_n = \sqrt{k^2 + \left(\frac{n\pi}{t}\right)^2}$, $F_{nn} = P_{nn} + \sin^2\theta \left[1 - P_{nn}(1 + \cos^2\varphi) + \frac{\omega_M}{\omega_H + \omega_M \lambda^2 k_n^2} P_{nn}(1 - P_{nn}) \sin^2\varphi \right]$, and $P_{nn} = \frac{k^2}{k_n^2} - \frac{k^4}{k_n^4}\frac{2}{kt}[1 - (-1)^n e^{-kt}]\frac{1}{1+\delta_{0n}}$. Under the experimental condition of a pure YIG film without additional material layers, a totally unpinned boundary condition is used. $n$ represents the $n^{\text{th}}$ standing wave mode in the thickness direction, where $n = 0$ is the uniform mode. The uniformity of the pumping magnetic field greatly suppresses the contribution of the $n > 0$ modes to the interaction Hamiltonian $H_{I,n} \propto \int_0^t dz\, h_{a.c.}(z) \cos\left(\frac{n\pi z}{t}\right)$, therefore we focused on $n = 0$ mode for quantitative calculations. For the experiments covered, $\theta = \frac{\pi}{2}$ and $\varphi$ represents the angle between the magnon wave vector ($\boldsymbol{k}$) direction and the applied static magnetic field $H$.

**Wave vector dependent coupling strength and pumping efficiency**

The coupling strength between the first resonator mode and the magnon mode with a wave vector $\boldsymbol{k}$ can be directly calculated by the average Zeeman interaction. Here we assume the resonator rf electromagnetic field always follows the magnon frequency and is quasi-TEM as commonly recognized for a microstrip geometry[31].

$$g_k \propto \int_{YIG} d^3r \, \langle \mathbf{h}_{rf}(\mathbf{r}; \omega_k) \cdot \mathbf{m}(\mathbf{r}; \mathbf{k}) \rangle \propto \int_{YIG} d^3r \, \left[ h_{y'}(y,z) - i e_k h_z(y,z) \right] \cos\left(\frac{\pi x}{L}\right) e^{-i\mathbf{k}\cdot\mathbf{r}}$$

Here $e_k = m_z/m_{y'}$ is the ratio between out-of-plane and in-plane magnetization amplitude of the magnon mode. For the uniform mode in a thin film, an estimation of $e_k \approx \sqrt{1 - \frac{4\pi M_S \cos^2\varphi}{H + Dk^2 + 4\pi M_S}}$ is used, considering both the surface demagnetization field and volume dipolar field. In the second integral, the first term in the bracket is the right-handed rf magnetic field that can effectively couple to the magnon mode; its $y, z$ dependency (in the cross-section plane) needs to be numerically calculated, while the cosine term represents its $x$ dependency (along the microstrip) which can be directly written. The last term represents the space distribution of magnon magnetization. Similarly, the pumping efficiency can be estimated by replacing the rf magnetic field with twice the magnon frequency and calculating the resulting average Zeeman interaction. Here this Zeeman interaction originates from magnon ellipticity $\rho_k = 1 - e_k^2 \approx \frac{4\pi M_S \cos^2\varphi}{H + Dk^2 + 4\pi M_S}$, which has to be explicitly included in the calculation. Crucially, because of momentum conservation, the two pumped magnons with opposite wave vector will form a standing wave, allowing parametric pumping for high wave vector magnons.

$$P_k \propto \int_{YIG} d^3r \, \langle \mathbf{h}_{rf}(\mathbf{r}; 2\omega_k) \cdot (\mathbf{m}(\mathbf{r}; \mathbf{k}) + \mathbf{m}(\mathbf{r}; -\mathbf{k})) \rangle$$

$$\propto \rho_k \int_{YIG} d^3r \, h_{x'}(y,z) \cos\left(\frac{2\pi x}{L}\right) \cos^2(\mathbf{k}\cdot\mathbf{r})$$

The $x'$ direction is parallel to the static external magnetic field and YIG magnetization, representing a parallel pumping configuration. Again, this parallel rf magnetic field component is numerically calculated from our device geometry.

**Transmission of the two-port microstrip resonator**

The FMR spectrum as a function of frequency $\omega$ and magnetic field $H$ can be fitted by the input-output theory for a cavity loaded with a magnet[20]:

$$s_{21}(\omega, H) = \frac{2\kappa_e}{2i(\omega - \omega_r) - \kappa_r + i\frac{4g}{\kappa_m - 2(\omega - \omega_m(H))}}$$

Here $\kappa_e$ is the external coupling rate between the resonator and the input/output ports; $\omega_r$ and $\kappa_r$ are the resonate frequency and dissipation rate; $\omega_m(H) = \gamma\sqrt{H(H + 4\pi M_S)}$ is the FMR frequency from Kittel formula; $\kappa_m$ is the FMR mode dissipation rate; $g$ is the coupling strength between FMR magnon and resonator photon; $\gamma$ is the gyromagnetic ratio; $M_S$ is the saturation magnetization of YIG film.

**Data availability**

Source data of this work are available from the corresponding authors upon reasonable request.

**Code availability**

Source codes used in this work are available from the corresponding authors upon reasonable request.

**Acknowledgements**

The research is primarily supported by National Science Foundation (NSF) under award ECCS-2309838. C.-T.C. acknowledges support from DOE-BES with award DE-SC0026239. A.K. is supported by NSF Research Experience for Undergraduate (REU) Program under award EEC-2349310.

**Author contributions**

L.L. and Q.W. conceived the project. Q.W. performed the experiments with help from C.-T.C. Q.W. analyzed the data with help from A.K. and L.L. Q.W, A.K and L.L wrote the paper. All authors discussed the experimental results.

**Competing interests**

Q.W. and L.L. are listed as inventors on a disclosure submitted to MIT Technology Licensing Office (TLO) about using coupled magnet-cavity system as correlated microwave photon source for communication and detection applications. MIT is pursuing patent protection based on this disclosure.

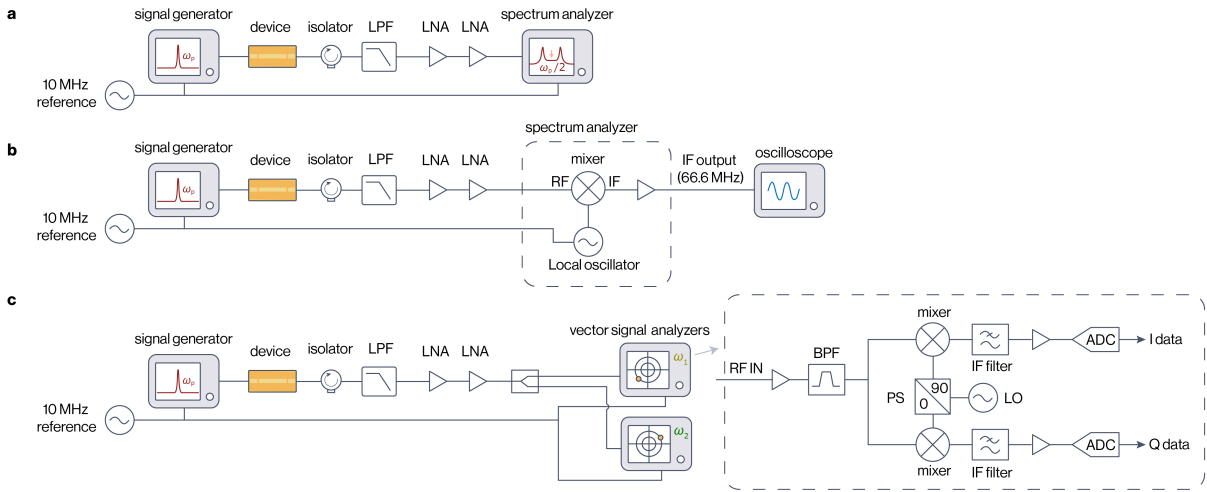

**Extended Fig. 1 | Measurement microwave circuits. a,** Circuit for spectrum measurements. **b,** Circuit for time-domain waveform measurement. **c,** Circuit for simultaneous quadrature measurement of two output signals with different angular frequency $\omega_1$ and $\omega_2$. LPF: low-pass filter; LNA: low-noise amplifier; RF: radio frequency; IF: intermediate frequency; LO: local oscillator; BPF: band-pass filter; PS: phase shifter; ADC: analog-to-digital converter. The components inside the dashed boxes are integrated RF units in the corresponding equipment.

# Supplementary Information for

# A room-temperature cavity-magnonic source of correlated microwave pairs

Qiuyuan Wang[1], Aravind Karthigeyan[1,2], Chung-Tao Chou[1,3], Luqiao Liu[1,*]

[1]Department of Electrical Engineering and Computer Science, Massachusetts Institute of Technology, Cambridge, MA, USA

[2]Department of Physics, University of Texas at Austin, Austin, Texas, USA

[3]Department of Physics, Massachusetts Institute of Technology, Cambridge, MA, USA

[*]Email: luqiao@mit.edu

**Supplementary Note 1. Theory for degenerate and non-degenerate parametric pumping in the cavity-magnonics system**

**1.1 Parametric oscillation modes**

Here we consider the case where parametric excitation is dominated by a pair of magnon modes, as we observed in our experiment. The minimum Hamiltonian, including a single pair of magnon modes $\hat{c}_{\pm k}$ with opposite wave vector $\pm k$, the resonator mode $\hat{a}_{r1}$ at $\omega_{r1}$, and the pumping microwave field $\hat{a}_{r2}$ at $\omega_{r2}$, is written as

$$\frac{\hat{H}}{\hbar} = \omega_{\pm k}\hat{c}^\dagger_{\pm k}\hat{c}_{\pm k} + \omega_{r1}\hat{a}^\dagger_{r1}\hat{a}_{r1} + \omega_{r2}\hat{a}^\dagger_{r2}\hat{a}_{r2} + g_{\pm k}\left(\hat{a}^\dagger_{r1}\hat{c}_{\pm k} + \hat{c}^\dagger_{\pm k}\hat{a}_{r1}\right)$$
$$+ P_k\left(\hat{a}^\dagger_{r2}\hat{c}_k\hat{c}_{-k} + \hat{c}^\dagger_k\hat{c}^\dagger_{-k}\hat{a}_{r2}\right) + \frac{\hat{H}_{4m}}{\hbar} \quad (1)$$

where the 4-magnon interaction term can be expressed as[1]

$$\frac{\hat{H}_{4m}}{\hbar} = \sum_{k,k'} \frac{1}{2} S_{k,k'} \hat{c}^\dagger_k \hat{c}^\dagger_{-k} \hat{c}_{k'} \hat{c}_{-k'} + T_{k,k'} \hat{c}^\dagger_k \hat{c}^\dagger_{k'} \hat{c}_k \hat{c}_{k'} \quad (2)$$

In Heisenberg picture, the corresponding equation of motion for the operator basis $\hat{V} = \left[\hat{c}_k, \hat{c}_{-k}, \hat{a}_{r1}, \hat{a}^\dagger_{r1}, \hat{c}^\dagger_k, \hat{c}^\dagger_{-k}\right]^T$ reads $\frac{d\hat{V}}{dt} = \frac{1}{i\hbar}[\hat{V}, \hat{H}] = M\hat{V}$. Under zero detuning ($\omega_{r1} = \omega_{\pm k} = \frac{1}{2}\omega_{r2}$) and a rotating frame with $\omega = \omega_{r1}$, we have (with magnon and cavity dissipation rate $\kappa_m$ and $\kappa_r$ introduced phenomenologically within $\omega_{\pm k}$ and $\omega_{r1}$)

$$M = \begin{bmatrix} -\kappa_m - iT\bar{n}_m & 0 & -ig & 0 & 0 & -i\alpha_p P_k \\ 0 & -\kappa_m - iT\bar{n}_m & -ig & 0 & -i\alpha_p P_k & 0 \\ -ig & -ig & -\kappa_r & 0 & 0 & 0 \\ 0 & 0 & 0 & -\kappa_r & ig & ig \\ 0 & i\alpha_p P_k & 0 & ig & -\kappa_m + iT\bar{n}_m & 0 \\ i\alpha_p P_k & 0 & 0 & ig & 0 & -\kappa_m + iT\bar{n}_m \end{bmatrix} \quad (3)$$

Here $\alpha_p = \langle \hat{a}_{r2} \rangle$ is the amplitude of the strong pumping field which is treated classically. Due to the symmetry between the $\pm k$ magnon modes, we simplify the equation of motion by writing $\bar{n}_m = \bar{n}_k = \bar{n}_{-k} = \langle \hat{c}_k^\dagger \hat{c}_k \rangle = \langle \hat{c}_{-k}^\dagger \hat{c}_{-k} \rangle$; $g = g_{-k} = g_k$; $T = 2S_{k,k'} = 4T_{k,k'} \approx \frac{4\omega_M}{NS}$ where $\omega_M = \gamma 4\pi M_S$; $N$ and $S$ are the total number of spins and spin quantum number, separately. Our experimental condition approximates $\kappa \approx \kappa_m \approx \kappa_r$, under which the eigen values of $M$ can be analytically expressed as

$$\lambda_{1,2} = -\kappa \pm \sqrt{(\alpha_p P_k)^2 - (T\bar{n}_m)^2} \tag{4.1}$$

$$\lambda_{3,4} = -\kappa + \frac{1}{2}\left( \sqrt{(\alpha_p P_k)^2 - (T\bar{n}_m)^2} \pm i\sqrt{8g^2 - ((\alpha_p P_k)^2 - (T\bar{n}_m)^2)} \right) \tag{4.2}$$

$$\lambda_{5,6} = -\kappa - \frac{1}{2}\left( \sqrt{(\alpha_p P_k)^2 - (T\bar{n}_m)^2} \pm i\sqrt{8g^2 - ((\alpha_p P_k)^2 - (T\bar{n}_m)^2)} \right) \tag{4.3}$$

Here $\lambda_{1,2}$ correspond to the eigen frequency of a dark magnon mode $\hat{c}_\pi = \frac{1}{\sqrt{2}}(\hat{c}_k - \hat{c}_{-k})$ that does not hybridize with $\hat{a}_{r1}$, therefore it cannot be detected by the output microwave signal. Here we notice that in the case without parametric pumping and four-magnon scattering ($\alpha_p = 0, \bar{n}_m = 0$), the rest eigen frequency is $\lambda = -\kappa \pm i\sqrt{2}g_k$. This is different from the well-known linear magnon polariton frequency[1] by a factor of $\sqrt{2}$. This factor comes from the fact that we considered three modes ($\hat{c}_k, \hat{c}_{-k}, \hat{a}_{r1}$), while in most calculations only two modes – the cavity and Kittel ($k = 0$) modes, are involved. To ensure the continuity under $k \to 0_\pm$, one can define $g_k' = \sqrt{2}g_k$. This $g_k'$ would correspond to the value measured in our experiment of Fig. 1g. The eigen values for magnon polariton modes are $-\kappa \pm \frac{1}{2}\left( \sqrt{(\alpha_p P_k)^2 - (T\bar{n}_m)^2} \pm i\sqrt{4g_k'^2 - ((\alpha_p P_k)^2 - (T\bar{n}_m)^2)} \right)$, with a maximum real part of $-\kappa + \frac{1}{2}\sqrt{(\alpha_p P_k)^2 - (T\bar{n}_m)^2}$. Under the steady state, we have this real part equal to zero, which gives us the saturation magnon number $\bar{n}_m = \frac{\sqrt{|\alpha_p P_k|^2 - 4\kappa^2}}{T}$. The eigen values are now

$$\lambda_{3,4} = \pm i\sqrt{g_k'^2 - \kappa^2} \tag{5.1}$$

$$\lambda_{5,6} = -2\kappa \pm i\sqrt{g_k'^2 - \kappa^2} \tag{5.2}$$

Below we only consider the modes corresponding to $\lambda_{3,4}$, as $\lambda_{5,6}$ has negative real part of the eigen frequency and will decay to zero quickly. When $g_k'$ is large, $\sqrt{g_k'^2 - \kappa^2}$ is real and $\lambda_{3,4}$ is purely imaginary, which leads to the steady state of non-degenerate parametric oscillation, with the split frequency $\Delta = \sqrt{g_k'^2 - \kappa^2}$. This can be realized for the case when the magnon polariton has low wave vector $\boldsymbol{k}$, which corresponds to high applied field. We notice that compared with the measured splitting $g'$ in the linear regime, the parametrically excited magnon polaritons have a much smaller splitting due to the close values between $g'$ and $\kappa$. For real experimental conditions, there will be a small detuning between the resonator and the magnon frequency, resulting in $\kappa$ being a weighted average of $\kappa_m$ and $\kappa_{r1}$, which could be even closer to $g'$.

While this expression provides a qualitative understanding on the observed degenerate and non-degenerate parametric oscillation, we note that this simplified model does not consider factors such as multiple magnon modes under different $\boldsymbol{k}$ directions, nonuniform magnetic field close to microstrip edges, or nonlinear damping when $\bar{n}_m$ is large, which will all contribute to the observed parametric excitations.

Setting $\bar{n}_m = 0$, we get the threshold pumping amplitude $\alpha_{th} P_k = 2\kappa$. For uniform pumping magnetic field[1], $\alpha P_{k,\text{uni}} = \gamma \rho_k h$, where $\rho_k \approx \frac{4\pi M_S \cos^2 \varphi}{H + Dk^2 + 4\pi M_S}$ (main text, methods) is the magnon ellipticity for a thin film sample, and $h$ is the rf field strength at YIG film location. When considering the finite geometry of real YIG sample, $P_k$ can be calculated numerically through $P_k \propto \rho_k \int_{\text{YIG}} d^3 r \, h_{x'}(y,z) \cos\left(\frac{2\pi x}{L}\right) \cos^2(\mathbf{k} \cdot \mathbf{r})$ (see Supplementary Fig. S1), with $x'$ being the direction of external dc magnetic field. Then, the threshold rf magnetic field can be estimated by $h_{th} \approx \frac{2\kappa}{\gamma \rho_k} \left(\frac{P_k}{P_{k,\text{uni}}}\right)^{-1}$, where the factor $\frac{P_k}{P_{k,\text{uni}}}$ can be numerically calculated based on the relationship For our device geometry, $h_{x'}(y,z)$ is approximately uniform, the $\cos\left(\frac{2\pi x}{L}\right)$ factor can also be replaced by constant within the YIG sample region. Therefore, we have the factor $\left(\frac{P_k}{P_{k,\text{uni}}}\right)^{-1} \approx 1$, and the threshold field $h_{th} \approx \frac{2\kappa}{\gamma \rho_k}$ is ~100 Oe, in reasonable agreement with the value extracted from the experimental threshold power, considering that the measured $\kappa_m$ contains inhomogeneity broadening contribution, and field distribution in the resonator is highly non-uniform.

## 1.2 Correlation of non-degenerate parametric oscillations

We analyze below the correlation and statistical properties of the polariton modes in the existence of thermal noise and 4-magnon interactions. First, in order to explicitly represent the magnon polaritons, we define polariton operator in the linear excitation region: $\hat{p}_\pm = u_k \hat{a}_{r1} + v_k \hat{c}_k + w_k \hat{c}_{-k}$, where $u_k$, $v_k$ and $w_k$ are coefficients determined by diagonalizing the $\mathbf{M}$ matrix (without $\alpha_p P_k$ and $T\bar{n}_m$ terms). In the case of zero-detuning, symmetric damping $\kappa \approx \kappa_m \approx \kappa_r$ and linear region, one has $u_k = \pm\frac{\sqrt{2}}{2}, v_k = w_k = \frac{1}{2}$. The corresponding magnon polariton Hamiltonian under rotating frame of $\omega_{r1}$ is $\frac{\hat{H}_{\text{pol}}}{\hbar} = \Omega \hat{p}_+^\dagger \hat{p}_+ - \Omega \hat{p}_-^\dagger \hat{p}_-$, where $\Omega = g'$ and we have dropped the dark modes. Now in the basis of (linear) magnon polaritons, we can rewrite the parametric pumping and four-magnon Hamiltonians

$$\frac{\hat{H}_{\text{para-int}}}{\hbar} = \sum_{i,j=+,-} P_{ij}\left(\hat{a}_{r2}^{\dagger}\hat{p}_i\hat{p}_j + \hat{p}_i^{\dagger}\hat{p}_j^{\dagger}\hat{a}_{r2}\right) \tag{6}$$

$$\frac{\hat{H}_{4m}}{\hbar} = \frac{1}{2}T\bar{n}\left(\hat{p}_+^{\dagger}\hat{p}_+ + \hat{p}_-^{\dagger}\hat{p}_-\right) \tag{7}$$

where $P_{+-} = P_{-+} = P_{++} = P_{--} = \frac{1}{2}P_k$; $\bar{n} = \bar{n}_+ = \bar{n}_- = \langle\hat{p}_+^{\dagger}\hat{p}_+\rangle = \langle\hat{p}_-^{\dagger}\hat{p}_-\rangle$. Switching to the polariton basis $\hat{Y} = [\hat{p}_+, \hat{p}_-, \hat{p}_+^{\dagger}, \hat{p}_-^{\dagger}]^T$, the equation of motion now has the following reduced form:

$$\frac{d\hat{Y}}{dt} = A\hat{Y} \text{ with } A = \begin{bmatrix} K_1 & L \\ L^* & K_2 \end{bmatrix}, \text{ where } K_1 = \begin{bmatrix} -i\Omega - \kappa - i\frac{1}{2}T\bar{n} & 0 \\ 0 & i\Omega - \kappa - i\frac{1}{2}T\bar{n} \end{bmatrix}, K_2 = \begin{bmatrix} i\Omega - \kappa + i\frac{1}{2}T\bar{n} & 0 \\ 0 & -i\Omega - \kappa + i\frac{1}{2}T\bar{n} \end{bmatrix}, \text{ and } L = -\frac{1}{2}i\alpha_p P_k \begin{bmatrix} 1 & 1 \\ 1 & 1 \end{bmatrix}.$$

The form of $\hat{H}_{\text{para-int}}$ suggests that parametric pumping process can excite a pair of polaritons with the same energy, e.g., $\hat{p}_+\hat{p}_+$, or polaritons with different ($\pm\Omega$) energies, e.g., $\hat{p}_+\hat{p}_-$. When exciting polaritons in the same energy branch, the summation of the energy from this pair of magnons has a finite detuning ($2\Omega$) from the pumping; therefore, this pumping process has a higher threshold pumping power and is suppressed. In the following, we focus only on the opposite branch parametric excitation. The system can be simplified by only considering the subspace $\hat{Z} = [\hat{p}_+, \hat{p}_-^{\dagger}]^T$. To include thermal noise, we add Hamiltonians corresponding to the thermal bath and the interaction between thermal bath and the magnon polariton system[2]:

$$\frac{\widehat{H}_R + \widehat{H}_{RS}}{\hbar} = \sum_l \omega_l \hat{R}_l^\dagger \hat{R}_l + \sum_{i=+,-}\sum_l \beta_{il}^* \hat{R}_l^\dagger \hat{p}_i + H.c.$$

with $\hat{R}_l^\dagger, \hat{R}_l$ being the operator for bath modes and $\beta_{il}$ being the interaction strength. The equation of motion in this further reduced space is:

$$\frac{d\widehat{\mathbf{Z}}}{dt} = \mathbf{N}\widehat{\mathbf{Z}} + \widehat{\mathbf{F}}(t) \tag{8}$$

with

$$\mathbf{N} = \begin{pmatrix} -i\Omega - \kappa - i\frac{1}{2}T\bar{n} & -i\frac{1}{2}\alpha_p P_k \\ i\frac{1}{2}\alpha_p P_k & -i\Omega - \kappa + i\frac{1}{2}T\bar{n} \end{pmatrix} \tag{9}$$

and

$$\widehat{\mathbf{F}}(t) = \begin{pmatrix} \hat{f}_+(t) \\ \hat{f}_-^\dagger(t) \end{pmatrix} = \begin{pmatrix} -i\sum_l \beta_{+l}\hat{R}_l \\ i\sum_l \beta_{-l}^* \hat{R}_l^\dagger \end{pmatrix} \tag{10}$$

Note that the Langevin force $\hat{f}_i(t)$ satisfies the following relation: $\langle \hat{f}_i(t) \rangle = 0$, $\langle \hat{f}_i(t)\hat{f}_i(t') \rangle = 0$, $\langle \hat{f}_i(t)\hat{f}_i^\dagger(t') \rangle = 2(\bar{n}_{i,th} + 1)\kappa\delta(t-t')$ and $\langle \hat{f}_i^\dagger(t)\hat{f}_i(t') \rangle = 2\bar{n}_{th}\kappa\delta(t-t')$, where $\bar{n}_{i,th}$ is the thermal polariton population. In general, the solution of the equation of motion above has the form of $\widehat{\mathbf{Z}}(t) = e^{\mathbf{N}t}\widehat{\mathbf{Z}}(0) + \int_0^t e^{\mathbf{N}(t-t')}\widehat{\mathbf{F}}(t')dt'$. We left multiply $\widehat{\mathbf{Z}}(t)$ with $\widehat{\mathbf{F}}^\dagger(t)$, use the assumption that $\langle \widehat{\mathbf{Z}}(0)\widehat{\mathbf{F}}^\dagger(t) \rangle = 0$ (no correlation between initial state of oscillator and noise), and plug in the correlation relationship of $\hat{f}_i(t)$, we get $\langle \hat{f}_i(t)\hat{p}_j(t) \rangle = 0$, $\langle \hat{f}_i^\dagger(t)\hat{p}_j^\dagger(t) \rangle = 0$, $\langle \hat{f}_i^\dagger(t)\hat{p}_i(t) \rangle = \langle \hat{f}_i(t)\hat{p}_i^\dagger(t) \rangle - 2\kappa = 2\bar{n}_{i,th}\kappa$. Next, we consider the equation of motion for $\hat{p}_+\hat{p}_-$. Using $\frac{d}{dt}(\hat{p}_+\hat{p}_-) = \frac{1}{i\hbar}[\hat{p}_+\hat{p}_-, \widehat{H}]$, we get:

$$\frac{d}{dt}\langle\hat{p}_+\hat{p}_-\rangle = (-2\kappa - iT\bar{n})\langle\hat{p}_+\hat{p}_-\rangle - i\left(\bar{n}+\frac{1}{2}\right)\alpha_p P_k - i\langle\hat{f}_+(t)\hat{p}_-(t)\rangle - i\langle\hat{f}_-(t)\hat{p}_+(t)\rangle \quad (11)$$

According to the paragraph above, the last two brackets are equal to zero, suggesting that $\langle\hat{p}_+\hat{p}_-\rangle$ is only driven by the effective force from the pumping field and thermal noise does not contribute (aside from indirectly through $\bar{n}$). For steady state, we have $\langle\hat{p}_+\hat{p}_-\rangle_{ss} = -\frac{i(2\bar{n}+1)\alpha_p P_k}{4\kappa+2iT\bar{n}}$, which has a fixed, time-independent phase. Applying $\hat{p}_+$ and $\hat{p}_-$ on polariton coherent states $|v_i\rangle$, we get $\hat{p}_+|v_+\rangle = v_+ e^{-i\Omega t}|v_+\rangle = r_+ e^{i\phi_+ - i\Omega t}|v_+\rangle$ and $\hat{p}_-|v_-\rangle = v_- e^{+i\Omega t}|v_-\rangle = r_- e^{i\phi_- + i\Omega t}|v_-\rangle$, where $v_\pm$ is the complex amplitude, $r_\pm$ and $\phi_\pm$ are defined as the real amplitude and phase, separately. The steady state solution of $\langle\hat{p}_+(t)\hat{p}_-(t)\rangle$ suggests that $\phi_+ + \phi_-$ is fixed, explaining the experimental observation.

### 1.3 Amplitude of oscillation

Solving Eq. (8) in the representation of coherent states, we get the corresponding equation of motion about the complex amplitude $v_+$ and $v_-^*$. Note that since we isolate the frequency difference of $\pm\Omega$ from the definition of $v_\pm$, the resulting equation of motion for the pair of non-degenerate oscillation complex amplitudes has the same form with the degenerate one[1], reproduced below (with necessary modifications):

$$\frac{d}{dt}v_\pm - \left[\frac{|\alpha_p P_k|^2 - 4\kappa^2}{8\kappa} - \frac{T^2}{8\kappa}\bar{n}^2\right]v_\pm = G_\pm(t) \quad (12)$$

with $G_\pm(t) = \frac{1}{2}f_\pm(t) + i\frac{\alpha_p P_k}{4\kappa}f_\mp^\dagger(t)$. Here approximation of $\frac{d^2}{dt^2}v_\pm \ll \frac{d}{dt}v_\pm$ is used, explaining the form of first-order equation rather than the expected second order one. The right-hand side of Eq. 12 is a small driving force related to thermal noise. Therefore, in order for a finite size steady state to exist, the bracket in front of the second term on the left-hand side should be equal to zero.

Otherwise, $v_{\pm}$ will either diverge or decay to zero, under the small thermal driving force. This gives the saturation polariton number $\bar{n} = \frac{\sqrt{|\alpha_p P_k|^2 - 4\kappa^2}}{T}$, the same result as that from section 1.1.

**1.4 Phase and oscillation linewidth**

The Langevin equation Eq. (12) can be transformed into a Fokker-Planck equation about $P$, which is the representation of the density matrix with the coherence state basis. $P(r_i, \phi_i, t)$ can be viewed as the probability distribution of the oscillator in the phase space defined by $r_i$ and $\phi_i$ ($i = \pm$):

$$\frac{d}{dt_n} P + \frac{1}{x}\frac{\partial}{\partial x}[(B - x^4)x^2 P] = \frac{1}{x}\frac{\partial}{\partial x}\left(x\frac{\partial P}{\partial x}\right) + \frac{1}{x^2}\frac{\partial^2 P}{\partial \phi_i^2} \tag{13}$$

with $t_n \equiv \left(\frac{T^2}{128}\kappa\bar{n}_{i,th}\right)^{\frac{1}{3}} t$, $x \equiv \left(\frac{T^2}{2\kappa^2 \bar{n}_{i,th}}\right)^{\frac{1}{6}} r_i$, and $B = 4\left(\frac{T^2}{16\kappa^2 \bar{n}_{i,th}}\right)^{\frac{2}{3}} \frac{|\alpha_p P_k|^2 - 4\kappa^2}{T^2}$. For steady state, $\frac{dP}{dt_n} = 0$, and $P$ is isotropic with angle $\phi$, i.e., $\frac{\partial P}{\partial \phi} = 0$, and only terms related to the derivative with respect to $x$ in Eq. (13) survive. The solution has the form $P(x) = Ne^{\frac{1}{2}Bx^2 - \frac{1}{6}x^6}$. When B is positive, the distribution function peaks at $x = B^{\frac{1}{4}}$, corresponding to $\bar{r}_i^2 = \frac{\sqrt{|\alpha_p P_k|^2 - 4\kappa^2}}{T}$, the same as $\bar{n}$ determined in section 1.1 and 1.3.

For large amplitude oscillation, $P(x)$ is narrowly peaked around $x_0 = \left(\frac{T^2}{2\kappa^2 \bar{n}_{i,th}}\right)^{\frac{1}{6}} \bar{r}_i$. When considering the evolution of probability distribution with time, we can assume that $P$ is fixed along the radial direction and only consider the change of $P$ with respect to $\phi$. Eq. (13) is simplified as:

$$\frac{d}{dt}P = D\frac{\partial^2 P}{\partial \phi_i^2} \tag{14}$$

with the diffusion coefficient $D = \frac{\bar{n}_{i,\text{th}}}{4\bar{n}}\kappa$. The solution of the one dimensional diffusion equation

Eq. (14) has the simple form of $P(\phi_i, t) = \frac{1}{\sqrt{4\pi D t}} e^{-\frac{\phi_i^2}{4Dt}}$. Let $v_+ = I_1 + iQ_1$ and $v_- = I_2 + iQ_2$ where I and Q are the in-phase and quadrature amplitudes of the polariton oscillator, we get the correlation function of $\langle I_1(t) I_1(t') \rangle = r_+^2 e^{-D|t-t'|} = \bar{n} e^{-D|t-t'|}$, representing the phase decoherence caused by phase diffusion[3]. $\langle I_1(t) I_1(t') \rangle$ shows that the auto-correlation from the quadrature measurement has the correlation time of $\tau = \frac{1}{D} = \frac{4\bar{n}}{\bar{n}_{i,\text{th}}} \frac{1}{\kappa}$. The Fourier transform of the auto-correlation function corresponds to the power spectrum density[4], which is a Lorentzian function with the linewidth given by $\Delta\omega_{\text{osc}} = D = \frac{\bar{n}_{i,\text{th}}}{4\bar{n}} \kappa \ll \kappa$, given the fact that under steady state, the thermal magnon polariton number is much smaller than the parametrically pumped number, $\bar{n}_{i,\text{th}} \ll \bar{n}$. In Supplementary Note 4, we calibrate $\bar{n}_{i,\text{th}}$ and $\bar{n}$ from the experimental measurement, which gives consistent results for $\Delta\omega_{\text{osc}}$ between this theory and experimental observation.

**Supplementary Note 2. Numerical results for magnon-photon coupling**

To quantify the wave vector ($k$) dependent coupling strength $g'_k$ and the parametric pumping efficiency $P_k$, we first numerically calculated the rf magnetic field distribution generated by the microstrip resonator, assuming a quasi-TEM mode[5]. The simulation considers the specific geometry of the 40 $\mu m$ narrowed microstrip and the YIG sample placement, as described in the main text and Methods. Supplementary Figure S1b shows the calculated rf field distribution in the $yz$ cross-sectional plane at the sample location. The in-plane field $h_y$ is found to be approximately uniform across the 3 $\mu m \times$ 40 $\mu m$ cross-section area, justifying the approximations used in Supplementary Section 1.1 for calculating $g'_k$ and $P_k$. Using these calculated rf field distributions and the expressions provided in the Methods section, we computed $g'_k$ and $P_k$. These calculations account for the external field angle ($\zeta = (90 - 38)°$) and assume the magnon wave vector $k$ is directed along the $y$ direction (perpendicular to the microstrip line, $\varphi = 38°$). Supplementary Figure S1c,d plots the calculated $g'_k$ and $P_k$ as a function of the magnon wave vector magnitude. We see that the normalized pumping efficiency $P_k$ (relative to the $k = 0$ mode) remains significant, varying between 0.5 and 1.0 over the relevant range of $k$. This is consistent with the parallel pumping configuration, which can efficiently excite standing waves formed by $\pm k$ magnons. In contrast, the coupling strength $g_k$ decreases rapidly as $k$ increases, approaching zero for large $k$. This reduction in $g'_k$ is attributed to the internal cancellation of the magnon mode's oscillating magnetization when averaged over the approximately uniform rf field profile. This $k$-dependence of $g'_k$ explains the transition from degenerate to non-degenerate pumping observed in main text Fig. 2, as $g'_k$ must be greater than the average dissipation $\kappa$ to open a frequency gap (see Supplementary Note 1, Eq. 5.1).

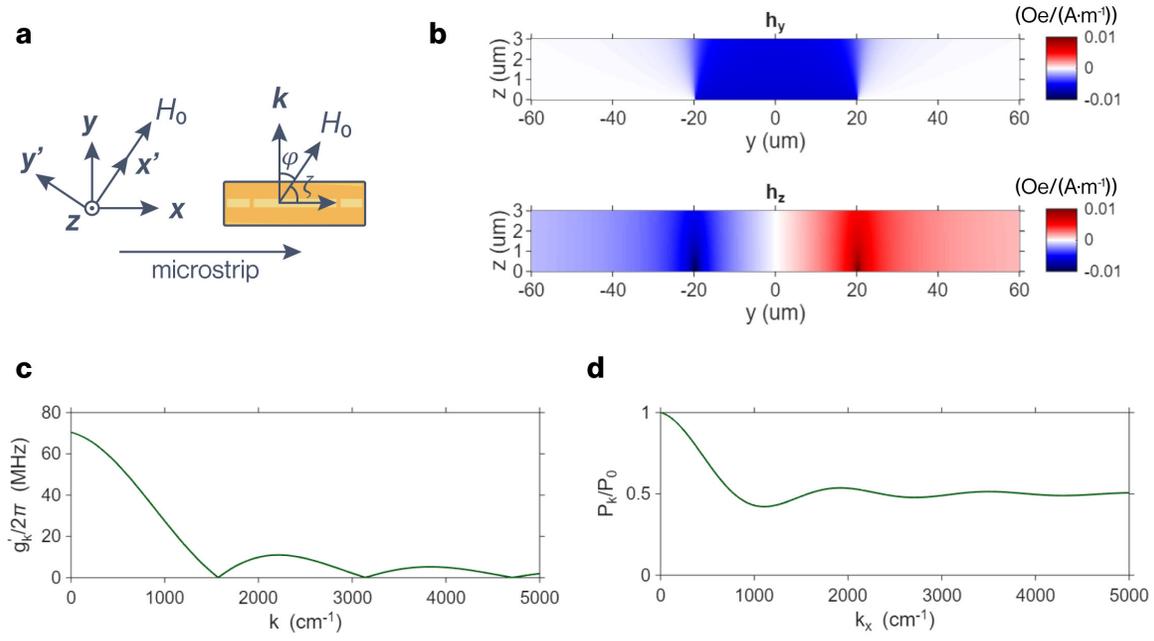

**Supplementary Figure S1.** RF field distribution across $yz$ plane. **a,** coordinate system used. **b,** calculated peak value of rf magnetic field components normalized by surface current density passing through the microstrip resonator. **c,d,** calculated coupling strength amplitude $|g'_k|$ (**c**) and normalized pumping efficiency $P_k/P_0$ under $\varphi = 38°$ and $\zeta = 52°$.

**Supplementary Note 3. Microwave resonator design**

The primary goal of the resonator design is to achieve the precise frequency-doubling relationship $\omega_{r2} = 2\omega_{r1}$, which is necessary for the parametric pumping scheme. For a standard, uniform microstrip resonator, the condition is naturally satisfied. However, the magnetic loading of YIG crystal shifts the permeability and the narrowed section complicates the harmonic relationship. The standing wave condition in such a width-varying resonator can be approximately expressed by treating the 40 $\mu m$ narrowed region as a transmission line with characteristic impedance $Z_1$:

$$\cot kl_1 + \alpha \cdot \frac{\cot kl_2 - \alpha \tan kl_0}{\alpha + \tan kl_0 \cot kl_2} = 0 \tag{15}$$

where $\alpha = Z_1/Z_0$, $Z_0$ is the characteristic impedance of the unnarrowed part, $l_1, l_2$ are the lengths of the unnarrowed region on left and right, respectively, and $l_0$ is the length of the narrowed region. Here we assume the change of effective dielectric constant is negligible as the transmission line width is much greater than conductor thickness of ~ 18 $\mu m$ and the microstrip resonator supports a quasi-TEM mode. The solutions of Eq. (15) are functions of ($l_1$, $l_2$, $l_0$), implying that the frequency ratio $\omega_{r2}(k_2)/\omega_{r1}(k_1)$ can be fine-tuned by varying the position and length of this narrowed region. To refine the design, we performed full-wave electromagnetic simulations using Sonnet software. These simulations allow precise calculation of the resonate frequencies and their corresponding quality factors (Q) and field distributions. Supplementary Figure S2 shows the simulated longitudinal current distributions for the half-wavelength ($\omega_{r1}$) and full-wavelength ($\omega_{r2}$) modes. This confirms that the YIG sample (positioned at ~ $\lambda_2/4$) is located near an antinode for both modes, which is crucial for maximizing both the magnon-photon coupling $g_k$ and the pumping efficiency $P_k$. The final resonator geometry was determined through an iterative process.

Several devices with slight variations in the length and position of the narrowed region were fabricated and tested. Based on these experimental measurements (e.g., the one shown in main text Fig. 1e), we finalized the geometry that yielded $\frac{\omega_{r1}}{2\pi}$ = 3.354 GHz and $\frac{\omega_{r2}}{2\pi}$ = 6.709 GHz, achieving the required frequency-doubling condition with high precision.

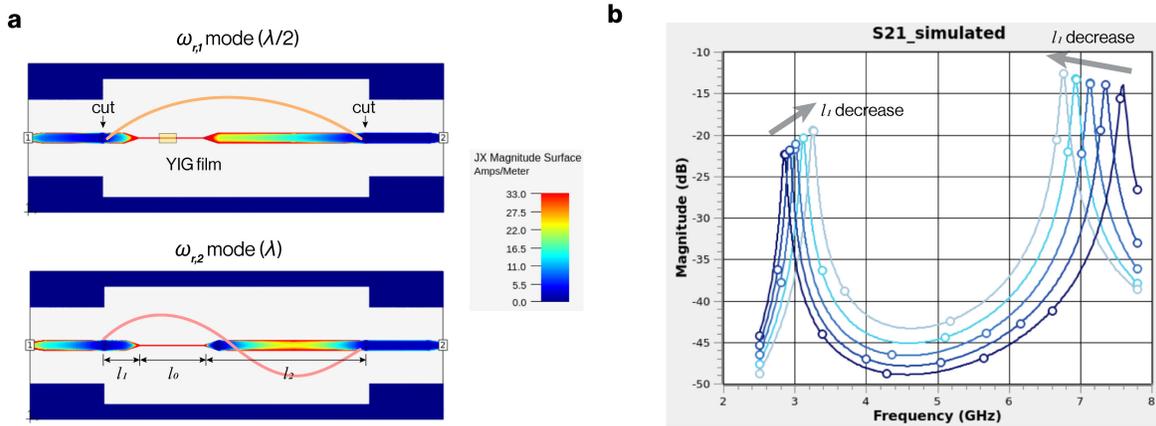

**Supplementary Figure S2. a,** Simulated rf field distribution for the first ($\omega_{r1}$) and second ($\omega_{r2}$) resonate mode of the microwave resonator. **b,** Simulated transmission parameter S21 under different resonator parameter $l_1$ while keeping $l_0$ and $l_1 + l_2 + l_0$ constant.

**Supplementary Note 4. Photon statistics of output modes**

The raw data acquired from the homodyne setup (Extended Fig. 1c) are quadrature voltages $(V_X(t), V_P(t))$. To analyze the polariton statistics, we first determine the conversion ratio between the measured voltage to the intracavity magnon polariton number by fitting the measured quadrature voltage distributions below the parametric pumping threshold ($P_{in}$ = 0 dBm) with a Boltzmann distribution $P_{th}(n) \propto e^{-n/\bar{n}_{th}}$, where $\bar{n}_{th} = \left(e^{\frac{\hbar\omega}{k_B T}} - 1\right)^{-1} \approx 1.86 \times 10^3$, as shown in Fig. S3a,b (data measured under 3.3532 GHz, corresponding to the lower frequency magnon polariton mode). Then, the measured quadrature data in main text fig. 2c (also shown here in Fig. S3c), is converted into a polariton number probability distribution $P(n)$. This distribution is then fitted using the steady-state solution derived from the Fokker-Planck equation in Supplementary Note 1:

$$P(n) = Ne^{\frac{1}{2}Bx^2 - \frac{1}{6}x^6} = Ne^{\frac{1}{2}B\beta^2 n - \frac{1}{6}\beta^6 n^3} \tag{16}$$

where $x \equiv \beta r = \left(\frac{T^2}{2\kappa^2 \bar{n}_{th}}\right)^{\frac{1}{6}} r$, $B = 4\left(\frac{T^2}{16\kappa^2 \bar{n}_{th}}\right)^{\frac{2}{3}} \frac{|\alpha_p P_k|^2 - 4\kappa^2}{T^2} = \beta^4 \frac{|\alpha_p P_k|^2 - 4\kappa^2}{T^2}$, and for large polariton number, $n = r^2$. The fitted curve is shown in Fig. S3d and we get the parameters $\beta \equiv \left(\frac{T^2}{2\kappa^2 \bar{n}_{th}}\right)^{\frac{1}{6}} \approx 7.95 \times 10^{-4}$, $B \approx 5.8828$, from which we can determine (1) the average steady-state magnon polariton number $\bar{n} \equiv \sqrt{\frac{B}{\beta^4}} \approx 3.835 \times 10^6$; (2) the linewidth of the output signal $\frac{\Delta\omega_{osc}}{2\pi} = \frac{D}{2\pi} = \frac{\bar{n}_{i,th}}{4\bar{n}} \frac{\kappa}{2\pi} \approx 1.215 \times 10^{-4} \times \frac{\kappa}{2\pi} = 8.36$ kHz and (3) the coherence time $\tau = \frac{2\pi}{\Delta\omega_{osc}} = 0.120$ ms. These parameters are in good agreement with experimental results.

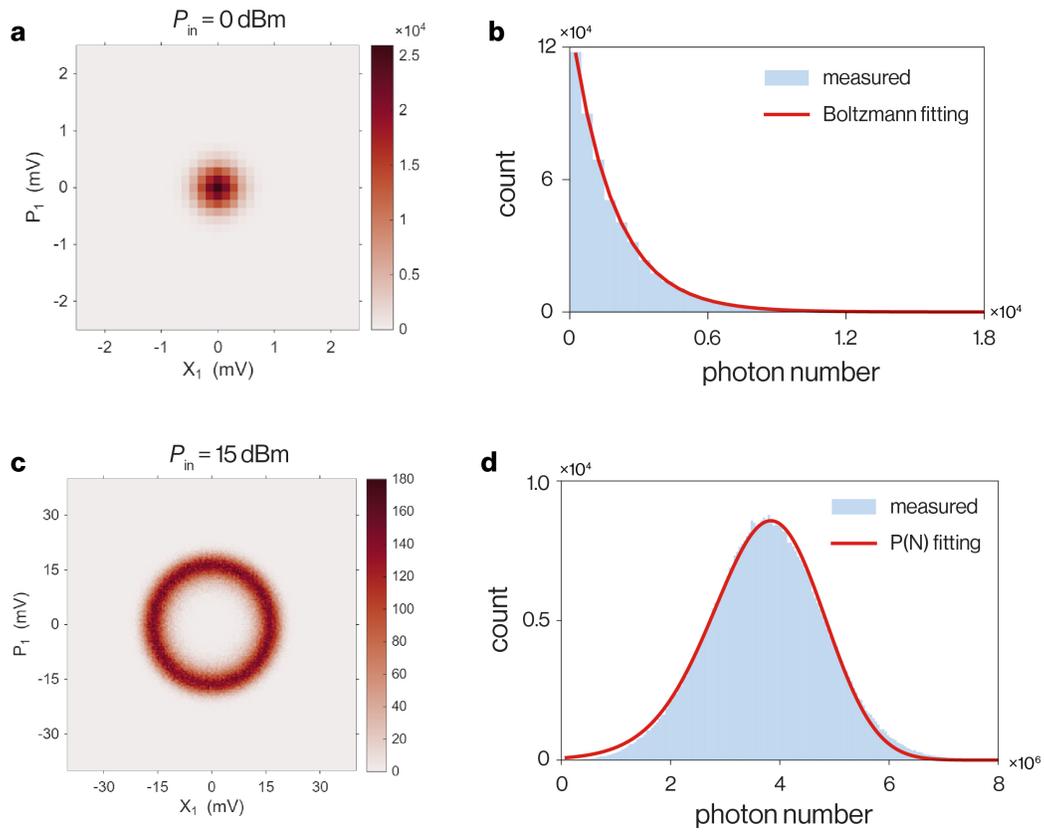

**Supplementary Figure S3.** Measured photon number distributions below (**a,b**) and above (**c,d**) the parametric pumping threshold.

**Supplementary Note 5. True random number generation test**

To evaluate the randomness of the output signal's phase, we processed the time-domain quadrature data measured for one of the non-degenerate modes (the 3.3532 GHz signal). The continuous phase $\varphi_1(t) = \tan^{-1}\frac{P_1(t)}{X_1(t)}$ was sampled at various time intervals from below and above the measured coherence time $\tau \approx 0.184$ ms. The sampled phase values were then digitized to generate a raw bitstream. This digitization was performed using several standard keying schemes, such as Binary Phase-Shift Keying (BPSK; mapping $[0, \pi)$ to '0' and $[\pi, 2\pi)$ to '1'), Quadrature Phase-Shift Keying (QPSK; 4-levels), and 8-Phase Shift Keying (8-PSK; 8-levels). The bitstreams generated were subsequently sent to the NIST SP800-22 statistical test suite for true randomness test. As noted in the main text (Fig. 3h), we excluded Maurer's Universal Statistical Test and the Approximate Entropy Test due to insufficient bitstream length for these specific sub-tests, leaving 14 applicable tests. Supplementary Table S1 summarizes results from each of the 14 NIST sub-tests for bitstreams generated with different sampling intervals and digitization schemes. As shown in the table and in Fig. 3h, the bitstreams generated with sampling intervals $\Delta T \geq 1.0$ ms successfully passed all 14 tests, confirming the true randomness from underlying thermal noise diffusion. In contrast, bitstreams sampled at lower $\Delta T$ failed more tests, as expected due to the short-term phase coherence. The two failed sub-tests for BPSK under 5.0 ms sampling interval is an artifact from insufficient bitstream length due to the short 5s measurement time.

| Digitization scheme | Sampling interval (ms) | 1 | 2 | 3 | 4 | 5 | 6 | 7 | 8 | 10 | 11 | 13 | 14 | 15 | 16 |
|---|---|---|---|---|---|---|---|---|---|---|---|---|---|---|---|
| BPSK | 5.0 | ✓ | ✓ | ✓ | ✓ | ✗ | ✓ | ✓ | ✗ | ✓ | ✓ | ✓ | ✓ | ✓ | ✓ |
| BPSK | 2.0 | ✓ | ✓ | ✓ | ✓ | ✓ | ✓ | ✓ | ✓ | ✓ | ✓ | ✓ | ✓ | ✓ | ✓ |
| BPSK | 1.0 | ✓ | ✓ | ✓ | ✓ | ✓ | ✓ | ✓ | ✓ | ✓ | ✓ | ✓ | ✓ | ✓ | ✓ |
| BPSK | 0.5 | ✓ | ✓ | ✗ | ✓ | ✓ | ✓ | ✓ | ✓ | ✓ | ✓ | ✓ | ✓ | ✓ | ✓ |
| BPSK | 0.2 | ✓ | ✗ | ✗ | ✗ | ✓ | ✗ | ✗ | ✗ | ✓ | ✗ | ✓ | ✓ | ✗ | ✓ |
| BPSK | 0.1 | ✓ | ✗ | ✗ | ✗ | ✓ | ✗ | ✗ | ✗ | ✗ | ✗ | ✗ | ✓ | ✗ | ✓ |
| QPSK | 5.0 | ✓ | ✓ | ✓ | ✓ | ✓ | ✓ | ✓ | ✓ | ✓ | ✓ | ✓ | ✓ | ✓ | ✓ |
| QPSK | 2.0 | ✓ | ✓ | ✓ | ✓ | ✓ | ✓ | ✓ | ✓ | ✓ | ✓ | ✓ | ✓ | ✓ | ✓ |
| QPSK | 1.0 | ✓ | ✓ | ✓ | ✓ | ✓ | ✓ | ✓ | ✓ | ✓ | ✓ | ✓ | ✓ | ✓ | ✓ |
| QPSK | 0.5 | ✓ | ✓ | ✗ | ✓ | ✓ | ✓ | ✓ | ✓ | ✓ | ✓ | ✓ | ✓ | ✓ | ✓ |
| QPSK | 0.2 | ✓ | ✗ | ✓ | ✗ | ✓ | ✗ | ✗ | ✗ | ✗ | ✓ | ✓ | ✓ | ✓ | ✓ |
| QPSK | 0.1 | ✓ | ✗ | ✗ | ✗ | ✓ | ✗ | ✗ | ✗ | ✗ | ✗ | ✗ | ✓ | ✓ | ✓ |
| 8-PSK | 5.0 | ✓ | ✓ | ✓ | ✓ | ✓ | ✓ | ✓ | ✓ | ✓ | ✓ | ✓ | ✓ | ✓ | ✓ |
| 8-PSK | 2.0 | ✓ | ✓ | ✓ | ✓ | ✓ | ✓ | ✓ | ✓ | ✓ | ✓ | ✓ | ✓ | ✓ | ✓ |
| 8-PSK | 1.0 | ✓ | ✓ | ✓ | ✓ | ✓ | ✓ | ✓ | ✓ | ✓ | ✓ | ✓ | ✓ | ✓ | ✓ |
| 8-PSK | 0.5 | ✓ | ✓ | ✗ | ✓ | ✓ | ✓ | ✓ | ✓ | ✓ | ✓ | ✓ | ✓ | ✓ | ✓ |
| 8-PSK | 0.2 | ✓ | ✗ | ✗ | ✓ | ✓ | ✓ | ✗ | ✗ | ✓ | ✗ | ✓ | ✓ | ✗ | ✓ |
| 8-PSK | 0.1 | ✓ | ✗ | ✗ | ✗ | ✓ | ✗ | ✗ | ✗ | ✗ | ✓ | ✓ | ✓ | ✓ | ✓ |

Column header: Sub test performed (green: passed, red: failed)

**Supplementary Table S1.** Summary of NIST SP800-22 statistical test results for bitstreams generated from the 3.3532 GHz signal. The 16 sub-tests, are referenced by ID: 1. Frequency (Monobit) test; 2. Frequency test within a block; 3. Runs test; 4. Test for the longest run of ones in a block; 5. Binary matrix rank test; 6. Discrete Fourier transform (spectral) test; 7. Non-overlapping template matching test; 8. Overlapping template matching test; 9. Maurer's "universal statistical" test; 10. Linear complexity test; 11. Serial test; 12. Approximate Entropy Test; 13. Cumulative sums test (forward); 14. Cumulative sums test (backward); 15. Random excursions test; 16. Random excursions variant test. Tests 9 and 12 are excluded as stated in the main text.

## Supplementary Note 6. Using correlated photon sources for RF communication with enhanced sensitivity and jamming resilience

Here we present enhanced sensitivity and jamming resilience by utilizing correlated microwave pairs as demonstrated in the main text. The transmitted signal can be expressed as $s_t(t) = Ae^{i[\omega_1 t + \varphi_1(t) + M(t)]}$, where $M(t)$ represents the information being coded using the IQ phase modulation. The received signal $s_r(t)$ consists of the attenuated transmitted signal $\alpha s_t(t)$ ($\alpha$ is the channel loss) plus a large, uncorrelated environmental noise component $s_n(t)$, such that $s_r(t) = \alpha s_t(t) + s_n(t)$. In a hostile or noisy environment, the noise power can be significantly larger than the signal power, resulting in a deeply negative signal-to-noise ratio (SNR). Traditional RF protocols would fail to recover the signal in this low-SNR regime. However, in our system, the receiver possesses the correlated *idler* channel, $s_i(t) = Be^{i[\omega_2 t + \varphi_2(t)]}$, which is strongly correlated with $s_t(t)$ but is uncorrelated with the noise $s_n(t)$. As depicted in main text Fig. 4a, we mix $s_r(t)$ with $s_i(t)$ and using a correlator, getting a signal

$$S(t;\tau) = s_r(t)s_i(t-\tau) = \{\alpha A e^{i[\omega_1 t + \varphi_1(t) + M(t)]} + s_n(t)\} B e^{i[\omega_2(t-\tau) + \varphi_2(t-\tau)]}$$

$$= \alpha A B e^{i[(\omega_1 + \omega_2)t + \varphi_1(t) + \varphi_2(t-\tau) - \omega_2 \tau + M(t)]} + s_n(t) B e^{i[\omega_2(t-\tau) + \varphi_2(t-\tau)]} \quad (17)$$

Here $\tau$ represents the time reference mismatch between transmitted signal and the idler. The first term is centered at the stable pumping frequency $\omega_p = \omega_1 + \omega_2$ and can be efficiently extracted after switching to rotating frame of $\omega_p$ and computes the cross-correlation $C(\tau)$ over an integration time $T_{\text{int}}$:

$$C(\tau) = \frac{1}{T_{\text{int}}} \int_0^{T_{\text{int}}} S(t;\tau) e^{-i\omega_p t} \, dt$$

$$= \frac{1}{T_{\text{int}}} \int_0^{T_{\text{int}}} \alpha A B e^{i[\varphi_1(t) + \varphi_2(t-\tau) - \omega_2 \tau + M(t)]} + s_n(t) B e^{i[-\omega_1 t - \omega_2 \tau + \varphi_2(t-\tau)]} \, dt \quad (18)$$

The first term in the integration will approach the original message $e^{iM(t)}$ when the correlator aligns the time frames so $\tau \to 0$. The second term is uncorrelated noise and will average to 0 as $T_{int}$ increases. The receiver can thus recover the transmitted data by monitoring the peak in $C(\tau)$, effectively 'pulling' the signal out of a noisy background. Further, trade-offs between communication security and efficiency can be made by adding uncorrelated noise bands in purpose and adjusting their relative strengths to the real signals.